# Second-Order Conductivity Probes a Cascade of Singularities in a Moiré Superlattice

Tanweer Ahmed,[1] Bao Q. Tu,[1,2] Kenji Watanabe,[3] Takashi Taniguchi,[4] Marco Gobbi,[5,6] Fèlix Casanova,[1,6] and Luis E. Hueso[1,6]

[1]CIC nanoGUNE BRTA, 20018 Donostia-San Sebastian, Basque Country, Spain.
[2]Departamento de Polímeros y Materiales Avanzados: Física, Química y Tecnología, University of the Basque Country (UPV/EHU), 20018 Donostia-San Sebastian, Basque Country, Spain
[3]Research Center for Electronic and Optical Materials, National Institute for Materials Science, 1-1 Namiki, Tsukuba 305-0044, Japan.
[4]Research Center for Materials Nanoarchitectonics, National Institute for Materials Science, 1-1 Namiki, Tsukuba 305-0044, Japan.
[5]Centro de Fısica de Materiales (CFM-MPC) Centro Mixto CSIC-UPV/EHU, 20018 Donostia-San Sebastian, Basque Country, Spain.
[6]IKERBASQUE, Basque Foundation for Science, 48009 Bilbao, Basque Country, Spain.

Systems lacking inversion symmetry inherently demonstrate a nonlinear electrical response (NLER) to an applied electric bias, emerging through extrinsic mechanisms. This response is highly sensitive to the electronic band structure, which can be engineered with remarkable precision in moiré superlattices formed from atomically thin quantum materials. Moiré superlattices host complex Fermi surface reconstructions near van Hove singularities (vHSs) in the electronic density of states. However, the role of these reconstructions in shaping NLER remains insufficiently understood. In this work, we systematically explore NLER in moiré superlattices of twisted double bilayer graphene (tDBLG) by tuning the Fermi level across multiple moiré bands on both sides of the charge neutrality point. We observe sharp variations and sign reversals in the NLER appearing via extrinsic pathways near mid-band vHSs. The second-order conductivity close to the vHSs demonstrates a much higher value than previous reports of extrinsic NLER in any other material. Our results demonstrate that NLER can serve as a sensitive probe of Fermi surface reconstructions and establish tDBLG as a versatile and highly efficient platform for generating and controlling the nonlinear electrical response.

The Fermi surface topology profoundly influences the electronic properties of materials [1], especially near the critical points [2] of electronic bands, where drastic changes to the topology/connectivity of the Fermi surface—Lifshitz transitions [3–6]—occur. These not only lead to a diverging density of states (DoS) or van Hove singularities (vHS) but also dramatically alter the velocity and effective mass of the charge carriers. While the Fermi energy ($E_F$) can be precisely tuned in atomically thin quantum materials, [7] vHSs in such systems are often located at electronically inaccessible energies ($E_F >$ 1 eV), as in single-layer (SLG) [8] and bilayer (BLG) [9] graphene, or at the band edge, as in BLG [10,11]. However, in the case of moiré superlattices, the vHSs can be tuned to appear at intermediate energies by choosing a suitable inter-layer twist angle [12–15]. Several exotic phenomena can co-exist near the vHSs in moiré superlattices. For example, a strong orbital angular moment in SLG-hexagonal boron nitride (hBN) [16], superconductivity in non-magic-angle SLG-SLG [17] and BLG-BLG moiré patterns [18], correlated [14,15,19], magnetically [20] ordered states, etc. Identifying and understanding vHSs are crucial to exploiting these novel states of matter appearing in artificial materials.

The vHSs in moiré superlattices can be identified using first-order electronic transport measurement only in the time-reversal symmetry broken condition by means of the Hall conductivity [14,15]. Recently, it has been demonstrated that the second-order nonlinear electrical response (NLER) [21–26] is highly sensitive to the electronic band structure [27–32] and to the Fermi surface topology [33,34] in the absence of time-reversal symmetry breaking. The NLER arises by virtue of the intrinsic Berry curvature dipole (BCD) [29,32,34,35] and/or by disorder-mediated extrinsic side-jump and skew scattering [24,27,28,36,37]. The extrinsic mechanisms only require inversion symmetry breaking and are strongly dependent on the effective mass and group velocity of the charge carriers [24,27], both of which show immense modulation as the $E_F$ is tuned across the vHSs. Multiple experimental observations of intrinsic and extrinsic NLER in moiré superlattices have been reported [27,28,31,32].

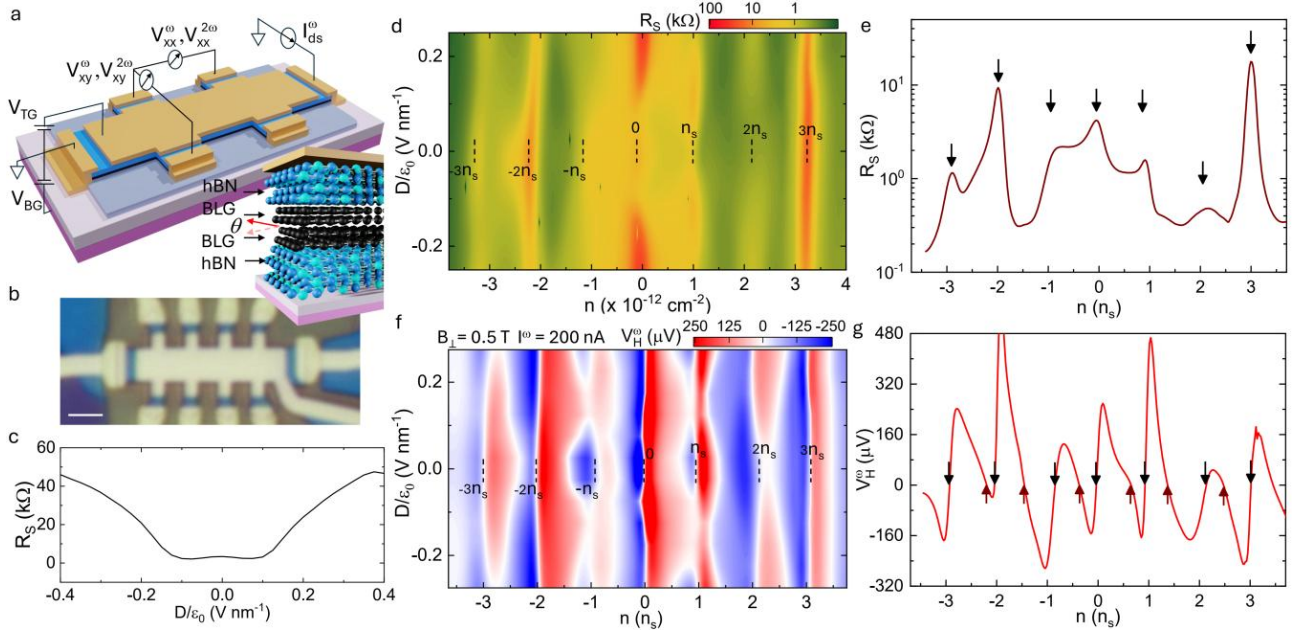

FIG. 1. **Device geometry and evidence of vHS in first-order electronic transport: a,** Schematic of the device and the measurement circuit. The inset shows the cross-section of the tDBLG channel. **b,** Optical image of the device with 2μm scalebar. **c,** $R_S$ vs $D$ characteristics at $n = 0$ indicating the topological change in the band structure. **d,** $n − D$ contour plot of $R_S$. The $R_S$ maxima at integer superlattice fillings are marked with vertical dashed lines. **e,** $R_S$ vs $n$ characteristics at $D = 0$. The integer fillings are marked with arrows. **f,** $n − D$ contour plot of $V_H^\omega$. **g,** $V_H^\omega$ vs $n$ data at $D = 0$. The sign changes close to the integer values of $n/n_s$ are marked with black arrows; the sign changes at non-integer values are marked with brown arrows.

However, the influence of vHSs on NLER, and whether NLER arising purely from extrinsic mechanisms can reliably detect vHSs, remain uncertain. Further experimental studies are needed to explore how NLER evolves near vHSs, particularly in engineered superlattices where multiple vHSs can be precisely tuned to appear within accessible Fermi energy ranges.

In this study, we explore NLER in ABAB-stacked twisted double bilayer graphene (tDBLG) moiré superlattices with an interlayer twist angle of $\theta \approx 0.7°$. By leveraging the small superlattice Brillouin zone at this twist angle, we modulate the Fermi level ($E_F$) up to fourth superlattice bands, *i.e.*, more than three times of the superlattice filling density ($n_s$) on both sides of the charge neutrality point (CNP). We identify the mid-band vHSs by measuring the Hall resistance as a function of the vertical displacement filed ($D$) and carrier concentration ($n$). We demonstrate that NLER changes signs close to all mid-band vHSs. Furthermore, the second-order conductivity ($\sigma^{2\omega}$) becomes pronounced close to the mid gap vHSs in the second and higher superlattice bands reaching a magnitude of 70 μmV$^{-1}$Ω$^{-1}$, a value one order of magnitude higher than any previous reports [27,28]. The temperature ($T$) dependent scaling behavior of the NLER close to the vHSs confirms that the observed second-order response is dominated by a combination of extrinsic-side jump and skew scattering mechanisms. Our study establishes NLER as a valuable tool for locating vHSs in artificial superlattices in time-reversal symmetric condition. Moreover, the mid-band vHSs at higher superlattice bands represent an efficient and tunable platform to generate NLER.

Encapsulated ABAB-stacked dual-gated tDBLG field effect transistor was fabricated using the usual tear and stack method [38] and dry transfer technique [39,40]. See methods and supporting information (SI) section 1 for fabrication and device details. Fig. 1a presents the device schematic and circuit diagram used for the simultaneous acquisition of first and second-harmonic longitudinal ($V_{xx}^\omega, V_{xx}^{2\omega}$) and transverse ($V_{xy}^\omega, V_{xy}^{2\omega}$) voltage drops. Inset schematically shows the cross-section of the channel; individual layers including the $\theta$ deg twisted BLG layers are marked. An optical image of the device with 2μm scalebar is shown in Fig. 1b. Fig. 1c shows the sheet resistance $R_S$ as a function of the vertical displacement field ($D$) measured at $T = 2K$, at $n = 0$. We tune $n$ and $D$ independently by simultaneously controlling the top ($V_{tg}$) and bottom ($V_{bg}$) gate voltages (see methods). See SI section 2 for mobility calculations. The electronic band structure in tDBLG can be substantially tuned by $D$ [38,41–43] unlike in tBLG moiré superlattices. The resistance sharply increases beyond $|D|/\varepsilon_0 > 0.15$ V nm$^{-1}$, marking a topological phase transition in the band structure, which is a characteristic feature of low-angle tDBLG. The $n - D$ contour plot of the $R_S$ presented in Fig. 1d further elucidates the electronic band structure of our tDBLG sample. We observe the charge neutrality point (CNP) at $n = 0$. Additionally, we observe resistance maxima at various positive and negative $n$. The high resistance states corresponding to integer band fillings appearing at $n \approx \pm 1.1, \pm 2.2, \pm 3.3 \times 10^{12}$ cm$^{-2}$, are marked as dashed vertical lines [42]. Clearly, these values correspond to integer multiples $n_s = 4/A \approx \pm 1.1 \times 10^{12}$ cm$^{-2}$, carrier concentration of one completely filled superlattice band. Here, $A \approx \sqrt{3}a^2/(2\theta^2)$ is the area of the moiré unit cell, and $a$ is graphene's lattice parameter [38,44]. From these expressions we calculate a twist angle $\theta \approx 0.92°$ (Importantly, the $R_s$ maxima consistently appear in the same positions throughout the channel, which excludes the possibility that a large twist-angle anomaly is responsible for the multiple peaks observed in $R_s$). For convenience, we describe the carrier density in units of $n_s$ in the following discussions. $R_s$ vs $n$ data at $D = 0$ V nm$^{-1}$ is presented in Fig. 1e. The $R_s$ peaks at integer values of $n/n_s$ are marked with black arrows. Also see SI section 4 for data from another sample (sample B) with a different $\theta \approx 1.44°$ twist angle.

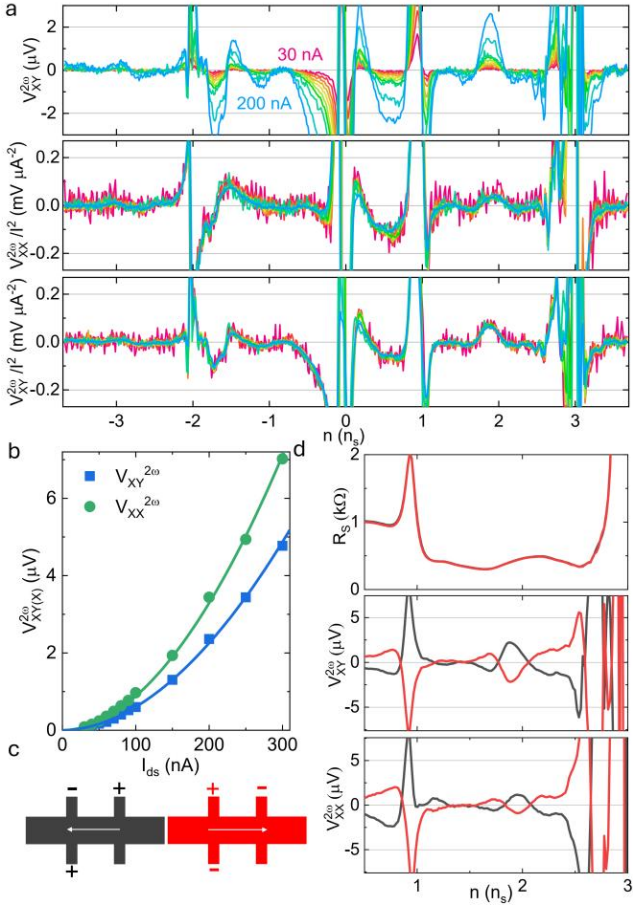

FIG. 2. **Second-order nonlinear transport in tDBLG: a,** The top panel shows $V_{xy}^{2\omega}$ vs $n$ data at different values of $I_{ds}^{\omega}$ at $D/\varepsilon_0 = -0.125$ Vnm$^{-1}$. The bottom (middle) panel shows normalized $V_{xy(x)}^{2\omega}/I_{ds}^2$ vs $n$ data at different $I_{ds}^{\omega}$, collapsing on each other. **b,** $|V_{xy(x)}^{2\omega}|$ vs $I_{ds}^{\omega}$ at $n/n_s \approx 0.5$ is shown as green (blue) data points. The solid traces are parabolic fits. **c,** Schematic of the two measurement configurations with switched polarities of voltage and current probes. **d,** The red and the black traces are the data at two measurement configurations. The top middle and bottom panels show the $n$-dependence of $R_s, V_{xy}^{2\omega}$, and $V_{xx}^{2\omega}$, respectively, at $D/\varepsilon_0 = -0.025$ Vnm$^{-1}$.

Fig. 1f presents the $n - D$ contour plot of the Hall voltage ($V_H^{\omega}$) at 500 mT magnetic field ($B_{\perp}$) with constant source-drain current $I_{ds}^{\omega} = 200$ nA. The data demonstrate abrupt sign reversals in $V_H^{\omega}$ at $n/n_s \approx 0, \pm 1, \pm 2, \pm 3$, indicating a change in the carrier type changes from hole to electron when $n/n_s$ is varied through the integer band fillings. The $V_H^{\omega}$ vs $n/n_s$ characteristic at $D = 0$ is presented in Fig. 1g. In addition to those at integer $n/n_s$, we also observe the sign reversals of $V_H^{\omega}$ at non-integer values of $n/n_s$, which can be attributed to the vHSs. The existence of the vHSs in the density of states of the SLG or BLG moiré superlattices are well known [12–15]. They are accompanied by the Lifshitz transitions where the topology/connectivity of the Fermi surface changes [45]. Such changes substantially modify the group velocity and effective mass of the charge carriers. Here, the nature of the charge carriers changes from electron-like to hole-like when the $E_F$ is increased through these reconstructions in the Fermi surface. Thus, the vHSs can be the detected using the $V_H^{\omega}$ [14]. The first order transport ($V_{xx}^{\omega}$) under time-reversal symmetric condition is insensitive to the vHSs, in our experimental conditions. However, when time-reversal symmetry is broken, i.e., in presence of vertical magnetic field, the vHSs lead to sign reversal in the first order transport ($V_H^{\omega}$) [14,15,18,45]. See SI section 4 for Hall data from sample B ($\theta \approx 1.44°$) showing tunable vHSs.

We now focus on the second order nonlinear electrical response (NLER) under time-reversal symmetry preserved condition [24,27]. NLER is a general property of any system with broken inversion symmetry and arises due to the non-zero elements of nonlinear conductivity tensor [36]. In the moiré systems, it is the interlayer twist that breaks the inversion symmetry and produces NLER, which has been previously reported on moiré superlattices based on SLG [27,28,32,33]. The NLER induces voltage drops at the second harmonic, or $2\omega$ frequency, either parallel ($V_{xx}^{2\omega}$) and/or perpendicular ($V_{xy}^{2\omega}$) to the applied current bias ($I_{ds}^{\omega}$). We present the $V_{xy}^{2\omega}$ vs $n$ data at $D/\varepsilon_0 = -0.125$ Vnm$^{-1}$ for different values of $I_{ds}^{\omega}$ ranging from 30 to 200 nA in the top panel of Fig. 2a. See SI section 3 for the $V_{xx}^{2\omega}$ data. The middle (bottom) panel of Fig. 2a shows the normalized $V_{xx(y)}^{2\omega}/I_{ds}^2$ vs $n$ data. Despite a low signal to noise ratio at low values of $I_{ds}^{\omega}$, normalized $V_{xx(y)}^{2\omega}/I_{ds}^{\omega\,2}$ data are observed to collapse on each other for the entire range of $n$, indicating a proportionality relationship $V_{xx(y)}^{2\omega} \propto I_{ds}^{\omega\,2}$. To further elucidate this, we present the $|V_{xy(x)}^{2\omega}|$ vs $I_{ds}^{\omega}$ data at $n \approx 0.5\,n_s$ as blue (green) data points, in Fig. 2b. The solid traces represent parabolic fits. Additionally, we measured the $V_{xy(x)}^{2\omega}$ in two different measurement configurations, reversing the polarity of the voltage and current probes [25]. The two configurations are schematically illustrated in Fig. 2c and highlighted in black and red. The polarity of the electrodes are indicated with + and − symbols. The top panel of Fig. 2d presents the $R_s - n$ data at $D/\varepsilon_0 = -0.025$ Vnm$^{-1}$ measured in the two configurations, showing a perfect overlap. The middle (bottom) panel presents the $V_{xy(x)}^{2\omega}$ vs $n$ data, which shows opposite signs for the two different measurement configurations (measured at $I_{ds}^{\omega} = 200$ nA). Notably both the $V_{xx}^{2\omega}$ and $V_{xy}^{2\omega}$ (i) show qualitatively and quantitatively similar with $n$, (ii) follow the proportionality relation $V_{xx(y)}^{2\omega} \propto I_{ds}^{\omega\,2}$. (iii) change sign when the measurement configuration is changed. These observations rule out any significant contribution of thermal effects in the measured $V_{xy(x)}^{2\omega}$ signal. Additionally, they also indicate the extrinsic mechanism arising from static and/or dynamic disorders as a primary contributor to the observed NLER across the entire $n$ range in our tDBLG sample [24,27,28].

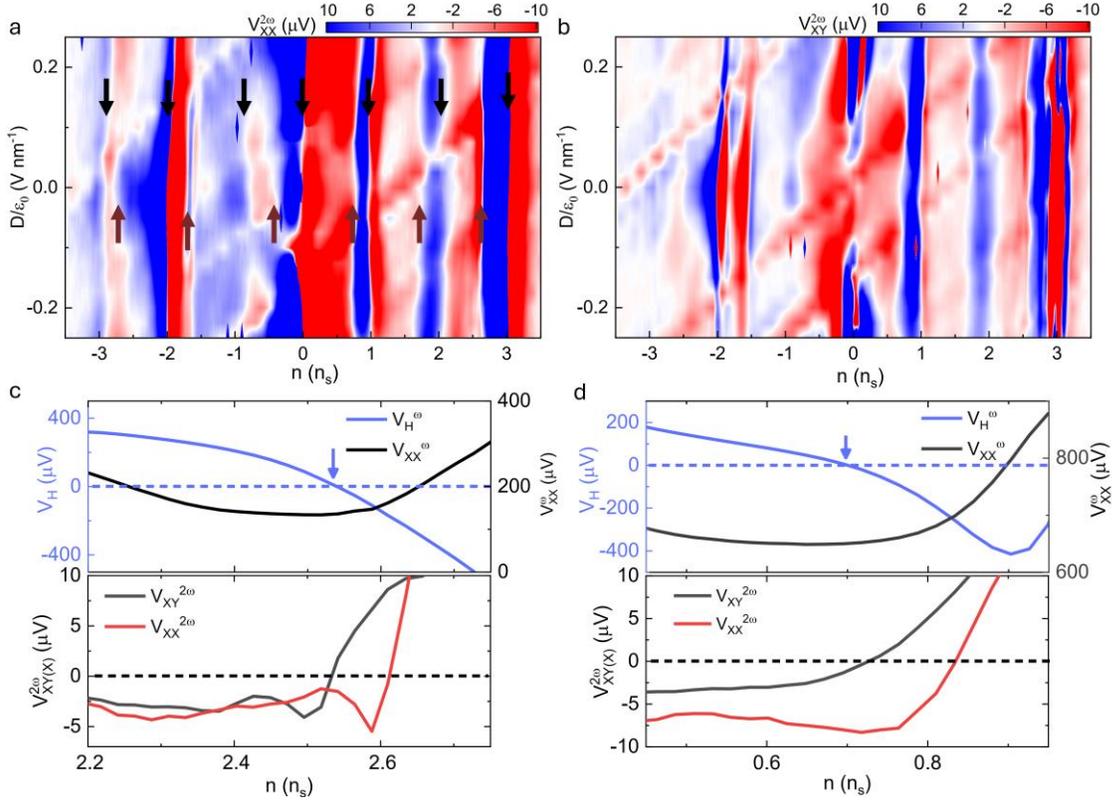

FIG. 3. $n-D$ **dependent evolution of NLER, observation of sign reversal near vHSs: a,** $n-D$ contour plot of $V_{xx}^{2\omega}$. The changes in $V_{xx}^{2\omega}$ at the integer and non-integer $n/n_s$ are indicated with black and brown arrows, respectively **b,** $n-D$ contour plot of $V_{xy}^{2\omega}$ is presented. **c,** The top (bottom) panel demonstrates $n$ dependence of $V_H^{\omega}$ and $R_s$ ($V_{xx}^{2\omega}$ and $V_{xy}^{2\omega}$ in time reversal symmetry preserved condition) at $D/\varepsilon_0 = 0.175$ Vnm$^{-1}$. **d,** The same data at $D/\varepsilon_0 = -0.025$ Vnm$^{-1}$ is presented. $V_H^{\omega}$ changes sign at the vHSs (indicated with an arrow). Both $V_{xx}^{2\omega}$ and $V_{xy}^{2\omega}$ changes sign in the vicinity of vHS.

$V_{xy(x)}^{2\omega}$ demonstrates strong variations as a function of $n$, indicating its sensitivity to the electronic structure of tDBLG. Interestingly, the sign change in the NLER is observed at $n/n_s \approx 0, \pm1, \pm2, \pm3$, reflecting the transition of the carriers from holes to electrons. Such sign changes

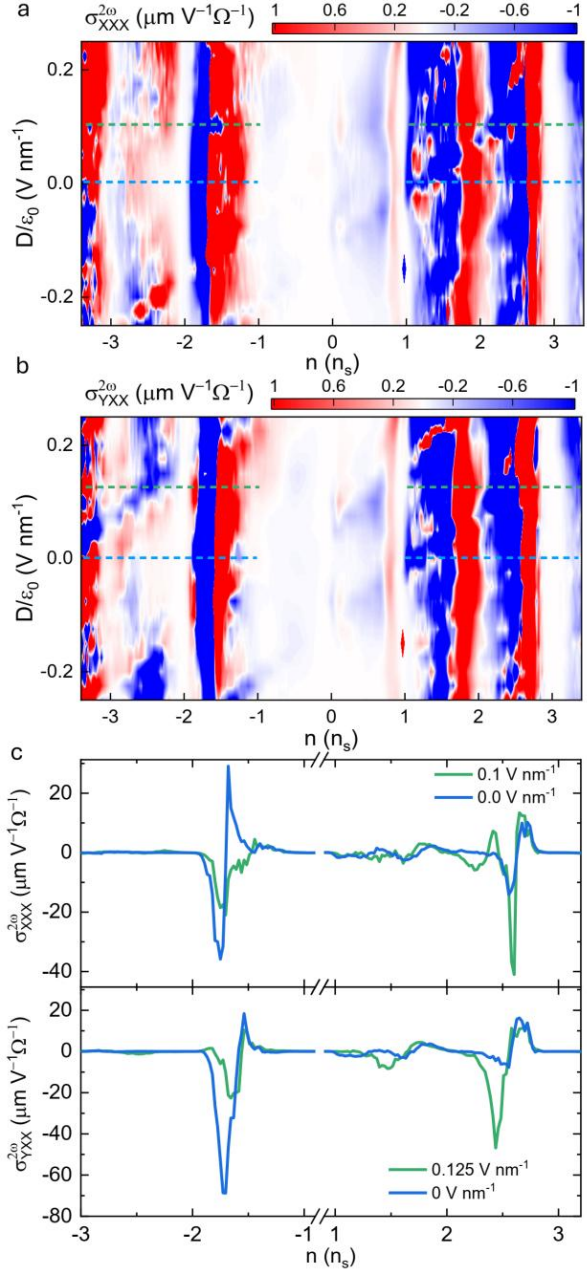

To elucidate the origin of these sign changes in the NLER at non-integer $n/n_s$ we investigate the $n-D$ contour plots of $V^{2\omega}_{xx(y)}$ which is presented in Fig. 3a (b). The diagonally dispersing features, appearing at fixed $V_{bg}$ are attributed to a contact effect arising from the non-top gated regions. The contour plots unambiguously demonstrate sign changes in the NLER in the vicinity of integer values of $n/n_s$, across all values of $D$, which is indicated with black arrows. The sign changes at non-integer $n/n_s$ are marked with brown arrows. The $n-D$ contour plots of $V^{2\omega}_{xy}$ and $V^{2\omega}_{xx}$ are observed to be qualitatively similar. The locus of the sign changes at the non-integer $n/n_s$ observed in $V^{2\omega}_{xx(y)}$ closely match those observed in and $V^{\omega}_H$ (Fig. 1f). To further highlight this, we compare $V^{2\omega}_{xx(y)}$ with $V^{\omega}_H$ and $R_s$ at a few different $D$ and $n$. The top (bottom) panel of Fig. 3c present the $n$-dependence of $V^{\omega}_H$ and $R_s$ ($V^{2\omega}_{xx(y)}$) at $D/\varepsilon_0 = 0.175$ V nm$^{-1}$ around $n/n_s = 2.5$. The location of the vHS giving rise to the sign change in $V^{\omega}_H$ is marked with the arrow $n \approx 2.5$. Notably, both $V^{2\omega}_{xx}$ and $V^{2\omega}_{xy}$ demonstrate sign changes in the vicinity of the vHS. Likewise, the $n$ dependences of $V^{\omega}_H, R_s, V^{2\omega}_{xx(y)}$ at $D/\varepsilon_0 = -0.025$ V nm$^{-1}$ (Fig. 3d) demonstrate the sign change of NLER close to the vHS at $n/n_s = 0.5$. The second-order conductivity [23,24] is significantly influenced by the electronic band structure [27,28,30–32] and the topology of the Fermi surface [17,34]. This is particularly important for extrinsic NLER assisted by impurity scattering, which is unavoidable at the Fermi surface. NLER is related to the Fermi distribution function's first and second derivatives [23], whose signs flip when the Fermi surface experiences a Lifshitz transition near the locations of vHSs. This is manifested as a sign reversal in the nonlinear electrical response (NLER) in the vicinity of the vHSs. Our experiment demonstrates that the NLER, under time-reversal symmetric conditions, is a sensitive probe for detecting mid-band vHSs emerging at non-integer values of $n/n_s$ in moiré superlattices. Also see SI section 5 for data from sample B ($\theta \approx 1.44°$) demonstrating the sign change in NLER close to vHSs.

Fig. 4a(b) shows the second-order longitudinal (transverse) conductivity extracted as $\sigma^{2\omega}_{xxx}(\sigma^{2\omega}_{yxx}) = \sigma V^{2\omega}_{xx}L/V^{\omega 2}_{xx}$ ($\sigma V^{2\omega}_{xy}L/V^{\omega 2}_{xx} \times L^2/W$) [27] as a function of $n$ and $D$. Here $L$, $W$ and $\sigma$ are the length, and width of the channel and first order conductivity, respectively (also see SI section 6). The second-order conductivity is observed to vary from a high negative to a high positive value in the vicinity of the mid-gap vHS. Interestingly, the values of both $\sigma^{2\omega}_{xxx}$ and $\sigma^{2\omega}_{yxx}$ have the same order of magnitudes, indicating they originate from the extrinsic mechanisms. We show the $\sigma^{2\omega}_{xxx}(\sigma^{2\omega}_{yxx})$ vs $n$ data at two different values at integer fillings have been previously observed in moiré superlattices of SLG-hBN [27] and tBLG. [28] The additional sign changes observed at non-integer values of $n/n_s$ in NLER data indicate the presence of finer features in the band structures that are difficult to observe in the first order transport under time-reversal symmetric conditions.

FIG.4. **Giant second-order conductivity near vHSs: a (b),** The $n-D$ contour plot of longitudinal (transverse) second order conductivity is presented. **c,** The top (bottom) panels present $\sigma^{2\omega}_{xxx}(\sigma^{2\omega}_{yxx})$ vs $n$ data at a few $D$, indicated as horizontal dashed lines. $\sigma^{2\omega}_{xxx}(\sigma^{2\omega}_{yxx})$ becomes pronounced close to the vHSs.

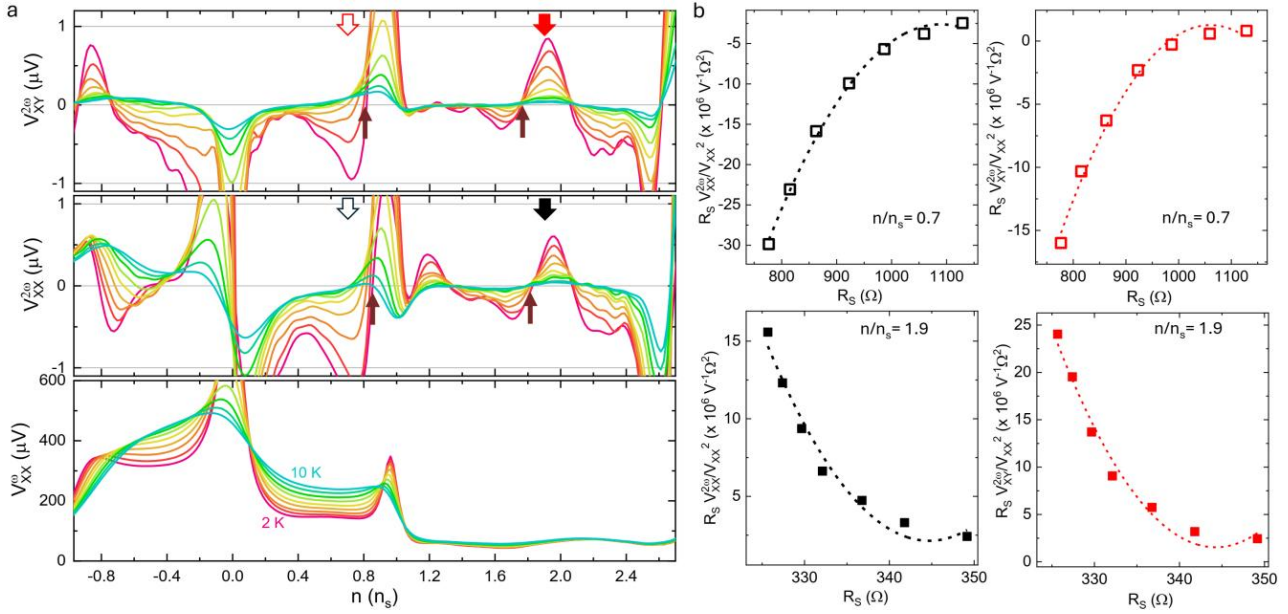

FIG. 5 . *T* dependent scaling behavior of NLER close to vHSs : **a,** The bottom panels presents the $R_s$ *vs* $n$ data at different T ranging from 2 to 10 K at $D = 0$. The top and middle panels present the $V_{xy}^{2\omega}$ and $V_{xx}^{2\omega}$ *vs* $n$ data at different temperatures. The brown arrows indicate the sign change in the $V_{xy}^{2\omega}$ and $V_{xx}^{2\omega}$ in the vicinity of the vHSs. **b,** The top left (right) panels presents $R_s^2 \, V_{xx(y)}^{2\omega}/V_{xx}^{\omega\,2}$ vs $R_s$ data at $n/n_s = 0.7$ marked with black (red) arrow. The similar data at $n/n_s = 1.9$ is presented in the bottom panels. The dotted lines indicate the fits using the generalized scaling behavior.

of $D$ in the top (bottom) panels of Fig. 4c. We observe giant second order conductivity in the second and higher superlattice bands. For example, at $D/\varepsilon_0 = 0$ V nm$^{-1}$, we achieve $|\sigma_{yxx}^{2\omega}| \approx 70$ µmV$^{-1}\Omega^{-1}$ near the mid-band vHS ($n/n_s \approx -1.7$) of the second superlattice valance band. The observed value is 10 times higher than any previously reported values [27,28]. A second order conductivity of this magnitude could pave the way for commercially viable and highly tunable rectifier devices [22,46].

The NLER in non-centrosymmetric crystals can appear from a combination of the extrinsic scattering mechanisms [24] and the intrinsic BCD dipole [21]. Extrinsic mechanisms lead to both transverse ($V_{xy}^{2\omega}$) and longitudinal ($V_{xx}^{2\omega}$) NLER. Pure intrinsic BCD, requiring additional C$_3$ symmetry breaking, leads to only $V_{xy}^{2\omega}$. To understand the origin of the observed NLER in our sample, we investigate its $T$-dependent scaling behavior. Here, we use $T$ as a tuning parameter for the scattering rates. In theory, the NLER is expected to follow a generalized scaling law given by

$$R_S^2 \frac{V_{xy(x)}^{2\omega}}{V_{xx}^{\omega\,2}} = \mathcal{C}_1 R_0^2 + \mathcal{C}_2 R_0 R_T + \mathcal{C}_3 R_T^2 \qquad (1)$$

Here, $R_0 = R_s(T \to 0)$, and $R_T = R_s(T) - R_0$. $\mathcal{C}_1$, $\mathcal{C}_2$, and $\mathcal{C}_3$ are the scaling parameters containing contributions from the intrinsic BCD (appearing only in $V_{xy}^{2\omega}$), side-jump and skew scattering from static and dynamic impurities [24]. See SI section 6 for more details. The bottom panel of Fig. 5a presents the $R_s - n$ data at different values of $T$, ranging from 2K to 10K at $D = 0$. The variation of $V_{xy(x)}^{2\omega}$ as a function of $n$, at the same values of $T$ is presented in the top (middle) panel of Fig. 5a. The brown arrows indicate the sign changes in the NLER at the non-integer values of $n/n_s$. We highlight the scaling behavior of the NLER at a few representative values of $n/n_s (\approx 0.7$ and $1.9$, indicated with black and red arrows), plotting the $R_s^2 \, V_{xy(x)}^{2\omega}/V_{xx}^{\omega\,2}$ vs $R_s$ characteristic in Fig. 5b as red (black) data points. The top and bottom panels present the data at $n/n_s = 0.7$ and $n/n_s = 1.9$, respectively (black arrows in Fig. 5a). The dotted lines present the fits to the data using equation 1. The black (red) fits in top panels of Fig. 5b, yields the following values of $\mathcal{C}_1$, $\mathcal{C}_2$, and $\mathcal{C}_3$, respectively: $-8.98 \times 10^{-5} \pm 2.65 \times 10^{-6}$ ($-5.16 \times 10^{-5} \pm 2.71 \times 10^{-6}$) V$^{-1}$, $3.05 \times 10^{-4} \pm 1.72 \times 10^{-5}$ ($2.11 \times 10^{-4} \pm 1.77 \times 10^{-5}$) V$^{-1}$ and $-2.74 \times 10^{-4} \pm 2.35 \times 10^{-5}$ ($-2.05 \times 10^{-4} \pm 2.43 \times 10^{-5}$) V$^{-1}$. All the parameters have the same order of magnitude, indicating the presence of the side-jump mechanism in addition to skew scattering from static and dynamic impurities. Likewise, from the black (red) fits in bottom panels of Fig. 5b, the following values of $\mathcal{C}_1$, $\mathcal{C}_2$, and $\mathcal{C}_3$ are obtained, respectively: $1.93 \times 10^{-4} \pm 1.2 \times 10^{-5}$ ($3.12 \times 10^{-4} \pm 1.85 \times 10^{-5}$) V$^{-1}$, $-4.89 \times 10^{-3} \pm 6.33 \times 10^{-4}$ ($-8.61 \times 10^{-3} \pm 9.5 \times 10^{-4}$) V$^{-1}$ and $3.47$

$\times 10^{-2} \pm 6.5 \times 10^{-3}$ ($6.23 \times 10^{-2} \pm 9.8 \times 10^{-3}$) V$^{-1}$. In this case, considering the large $\mathcal{C}_3$, skew scattering is identified as the dominant mechanism.

In conclusion, we performed simultaneous first and second order conductivity measurements on a moiré superlattice of twisted double bilayer graphene (tDBLG) by varying the Fermi energy across four superlattice bands on both sides of the CNP. We identified a cascade of mid-band singularities by performing Hall measurements at low magnetic fields. We demonstrate that under time-reversal symmetric condition, the nonlinear conductivity arising from a combination of extrinsic side-jump and skew scattering mechanism, changes sign while exibiting local extrema close to the vHSs, over the whole range of the superlattice bands. Our study establishes the second-order conductivity as a valuable tool to identify vHSs under time-reversal symmetric preserved condition. We demonstrate giant second-order conductivities close to the vHSs of second and higher superlattice bands. The observed magnitude $\approx$ 70 μmV$^{-1}\Omega^{-1}$ is 10 times higher than that of any other materials at similar thermodynamic conditions. Our work cements vHSs in tDBLG to be the most efficient and tunable platform to generate second-order nonlinear electrical response.

**METHODS:**
**Device fabrication:** The hBN and BLG crystals were individually exfoliated onto clean SiO$_2$ (300 nm ± 5%) / p++-Si substrates. hBN flakes with thicknesses ranging from 15 to 40 nm were selected as encapsulation and dielectric layers. The heterostructures were assembled using the dry transfer method, employing the pickup and stack technique within an Ar-filled glovebox equipped with robotic manipulators. The BLG flake was mechanically torn using a sharp tip, for creating tDBLG superlattice. Detailed steps for sample fabrication, including the tear-and-stack process, are provided in SI Section 1.

**Electrical measurements:** Hexagonal boron nitride (hBN) was used as the top-gate dielectric, with a 300 nm thick layer of SiO$_2$ serving as the bottom-gate dielectric. The carrier density is $n$ defined as $n = (\epsilon_0/e)[(\epsilon_{tg}(V_{tg} - V_{t0})/d_{tg}) + (\epsilon_{bg}(V_{bg} - V_{b0})/d_{bg})]$, where $V_{tg}$ and $V_{bg}$ represent the top and bottom gate voltages respectively. By simultaneously adjusting $V_{tg}$ and $V_{bg}$ while maintaining a constant displacement field $D$ calculated using the expression $2D/\epsilon_0 = [(\epsilon_{tg}(V_{tg} - V_{t0})/d_{tg}) - (\epsilon_{bg}(V_{bg} - V_{b0})/d_{bg})]$, where, $e$ is the electronic charge, $\epsilon_0$ is the vacuum permittivity, and $\epsilon_{tg}$ ($d_{tg}$) and $\epsilon_{bg}$ ($d_{bg}$) indicate the dielectric constants (thicknesses) of the top and bottom gate dielectrics, respectively; $V_{b0}$ and $V_{t0}$ represent the gate voltage offsets required to reach the charge neutrality point (CNP). The thickness of the top-gate dielectric ($d_{tg}$) was determined from the progression of the CNP within the $V_{tg}$-$V_{bg}$ phase space of sheet resistance ($R_S$), and this result was cross-verified through Hall measurements. Electrical measurements were performed using synchronized lock-in amplifiers with reference frequencies of 77.77 Hz and 17.77 Hz, both having input impedances significantly higher than $R_S$. No significant difference was observed between the two frequencies. The lock-in reference signal output was transformed into a current source using a high series-resistance ($\gg R_S$). Dual-channel DC source meters were employed to apply the gate voltages. The measurements were conducted within a closed-cycle cryostat, spanning a temperature range from 1.8 K to 300 K, with a vertical magnetic field range of ±9 T.

**Supporting information:** The Supporting information (SI) is available for free from the publisher´s website. SI contains the following sections: 1. Device fabrication method, and sample details. 2. Calculation of field-effect and Hall mobility. 3. $V_{xx}^{2\omega}$ vs n at different values of current bias ($I_{ds}^\omega$). 4. Evidence of mid-gap vHSs from sample B. 5. Sign change of NLER close to vHSs from sample B. 6. Scaling theory of NLER.

**Data availability statement:** All data needed to evaluate the conclusions in the paper are included in the paper and/or the supporting information. All source data related to the findings in this study can be obtained from the authors. All the experimental data will be uploaded to the Zenodo repository upon publication.

**Author contributions:** T.A. and L.E.H. conceived the project. T.A. fabricated the samples, performed the measurements, analyzed the data and wrote the manuscript. B.Q.T. assisted in the experiments. T.T. and K.W. provided the hBN crystals. All authors discussed, commented on the paper. note: The authors declare that they have no competing interests.

**ACKNOWLEDGEMENTS:** The authors acknowledge financial support from MICIU/AEI/10.13039/ 501100011033 (Grant CEX2020-001038-M), from MICIU/AEI and ERDF/EU (Projects PID2021-122511OB-I00 and PID2021-128004NB-C21), from MICIU/AEI and European Union NextGenerationEU /PRTR (Grant PCI2021-122038-2A) and from the European Union (Project 101046231-FantastiCOF). T.A. acknowledges funding from the European Union under Marie Sklodowska-Curie grant agreement number 101107842 (ACCESS). K.W. and T.T. acknowledge support from the JSPS KAKENHI (Grant Numbers 21H05233 and 23H02052) and World Premier International Research Center Initiative (WPI), MEXT, Japan.

**References:**

[1] I. M. Lifshitz, Anomalies of electron characteristics of a metal in the high pressure region, Sov. Phys. JEPT **11**, 1130 (1960).

[2] L. Van Hove, The Occurrence of Singularities in the Elastic Frequency Distribution of a Crystal, Phys. Rev. **89**, 1189 (1953).

[3] G. Bastien, A. Gourgout, D. Aoki, A. Pourret, I. Sheikin, G. Seyfarth, J. Flouquet, and G. Knebel, Lifshitz Transitions in the Ferromagnetic Superconductor UCoGe, Phys. Rev. Lett. **117**, 206401 (2016).

[4] X. Shi et al., Enhanced superconductivity accompanying a Lifshitz transition in electron-doped FeSe monolayer, Nat. Commun. **8**, 14988 (2017).

[5] C. Liu et al., Evidence for a Lifshitz transition in electron-doped iron arsenic superconductors at the onset of superconductivity, Nat. Phys. **6**, 419 (2010).

[6] H. Zhou, L. Holleis, Y. Saito, L. Cohen, W. Huynh, C. L. Patterson, F. Yang, T. Taniguchi, K. Watanabe, and A. F. Young, Isospin magnetism and spin-polarized superconductivity in Bernal bilayer graphene, Science **375**, 774 (2022).

[7] A. K. Geim and K. S. Novoselov, The rise of graphene, Nat. Mater. **6**, 183 (2007).

[8] A. H. Castro Neto, F. Guinea, N. M. R. Peres, K. S. Novoselov, and A. K. Geim, The electronic properties of graphene, Rev. Mod. Phys. **81**, 109 (2009).

[9] E. McCann and M. Koshino, The electronic properties of bilayer graphene, Rep. Prog. Phys. **76**, 056503 (2013).

[10] A. Varlet, D. Bischoff, P. Simonet, K. Watanabe, T. Taniguchi, T. Ihn, K. Ensslin, M. Mucha-Kruczyński, and V. I. Fal'ko, Anomalous Sequence of Quantum Hall Liquids Revealing a Tunable Lifshitz Transition in Bilayer Graphene, Phys. Rev. Lett. **113**, 116602 (2014).

[11] A. M. Seiler, N. Jacobsen, M. Statz, N. Fernandez, F. Falorsi, K. Watanabe, T. Taniguchi, Z. Dong, L. S. Levitov, and R. T. Weitz, Probing the tunable multi-cone band structure in Bernal bilayer graphene, Nat. Commun. **15**, 3133 (2024).

[12] I. Brihuega, P. Mallet, H. González-Herrero, G. Trambly de Laissardière, M. M. Ugeda, L. Magaud, J. M. Gómez-Rodríguez, F. Ynduráin, and J.-Y. Veuillen, Unraveling the Intrinsic and Robust Nature of van Hove Singularities in Twisted Bilayer Graphene by Scanning Tunneling Microscopy and Theoretical Analysis, Phys. Rev. Lett. **109**, 196802 (2012).

[13] G. Li, A. Luican, J. M. B. Lopes dos Santos, A. H. Castro Neto, A. Reina, J. Kong, and E. Y. Andrei, Observation of Van Hove singularities in twisted graphene layers, Nat. Phys. **6**, 109 (2010).

[14] S. Wu, Z. Zhang, K. Watanabe, T. Taniguchi, and E. Y. Andrei, Chern insulators, van Hove singularities and topological flat bands in magic-angle twisted bilayer graphene, Nat. Mater. **20**, 488 (2021).

[15] S. Xu et al., Tunable van Hove singularities and correlated states in twisted monolayer–bilayer graphene, Nat. Phys. **17**, 619 (2021).

[16] R. Moriya, K. Kinoshita, J. A. Crosse, K. Watanabe, T. Taniguchi, S. Masubuchi, P. Moon, M. Koshino, and T. Machida, Emergence of orbital angular moment at van Hove singularity in graphene/h-BN moiré superlattice, Nat. Commun. **11**, 5380 (2020).

[17] R. Dutta, A. Ghosh, S. Mandal, K. Watanabe, T. Taniguchi, H. R. Krishnamurthy, S. Banerjee, M. Jain, and A. Das, *Electric Field Tunable Superconductivity with Competing Orders in Twisted Bilayer Graphene near Magic-Angle*, arXiv:2402.11649.

[18] R. Su, M. Kuiri, K. Watanabe, T. Taniguchi, and J. Folk, Superconductivity in twisted double bilayer graphene stabilized by WSe2, Nat. Mater. **22**, 1332 (2023).

[19] P. Knüppel, J. Zhu, Y. Xia, Z. Xia, Z. Han, Y. Zeng, K. Watanabe, T. Taniguchi, J. Shan, and K. F. Mak, *Correlated States Controlled by Tunable van Hove Singularity in Moiré WSe2*, arXiv:2406.03315.

[20] Y.-W. Liu, J.-B. Qiao, C. Yan, Y. Zhang, S.-Y. Li, and L. He, Magnetism near half-filling of a Van Hove singularity in twisted graphene bilayer, Phys. Rev. B **99**, 201408 (2019).

[21] I. Sodemann, Quantum Nonlinear Hall Effect Induced by Berry Curvature Dipole in Time-Reversal Invariant Materials, Phys. Rev. Lett. **115**, (2015).

[22] H. Isobe, S.-Y. Xu, and L. Fu, High-frequency rectification via chiral Bloch electrons, Sci. Adv. **6**, eaay2497 (2020).

[23] Z. Z. Du, H.-Z. Lu, and X. C. Xie, Nonlinear Hall effects, Nat. Rev. Phys. **3**, 744 (2021).

[24] Z. Z. Du, C. M. Wang, S. Li, H.-Z. Lu, and X. C. Xie, Disorder-induced nonlinear Hall effect with time-reversal symmetry, Nat. Commun. **10**, 3047 (2019).

[25] M. Ramos et al., Unveiling Intrinsic Bulk Photovoltaic Effect in Atomically Thin ReS2, Nano Lett. **24**, 14728 (2024).

[26] T. Ahmed, H. Varshney, B. Q. Tu, K. Watanabe, T. Taniguchi, M. Gobbi, F. Casanova, A. Agarwal, and L. E. Hueso, Detecting Lifshitz Transitions Using Nonlinear Conductivity in Bilayer Graphene, Small **21**, 2501426 (2025).

[27] P. He, G. K. W. Koon, H. Isobe, J. Y. Tan, J. Hu, A. H. C. Neto, L. Fu, and H. Yang, Graphene moiré superlattices with giant quantum nonlinearity of chiral Bloch electrons, Nat. Nanotechnol. **17**, 378 (2022).

[28] J. Duan, Y. Jian, Y. Gao, H. Peng, J. Zhong, Q. Feng, J. Mao, and Y. Yao, Giant Second-Order Nonlinear Hall Effect in Twisted Bilayer Graphene, Phys. Rev. Lett. **129**, 186801 (2022).

[29] Q. Ma et al., Observation of the nonlinear Hall effect under time-reversal-symmetric conditions, Nature **565**, 337 (2019).

[30] Z. He and H. Weng, Giant nonlinear Hall effect in twisted bilayer WTe2, Npj Quantum Mater. **6**, 1 (2021).

[31] M. Huang et al., Giant nonlinear Hall effect in twisted bilayer WSe2, Natl. Sci. Rev. **10**, nwac232 (2023).

[32] M. Huang et al., Intrinsic Nonlinear Hall Effect and Gate-Switchable Berry Curvature Sliding in Twisted Bilayer Graphene, Phys. Rev. Lett. **131**, 066301 (2023).

[33] S. Datta, S. Bhowmik, H. Varshney, K. Watanabe, T. Taniguchi, A. Agarwal, and U. Chandni, Nonlinear Electrical Transport Unveils Fermi Surface Malleability in a Moiré Heterostructure, Nano Lett. **24**, 9520 (2024).

[34] S. Sinha et al., Berry curvature dipole senses topological transition in a moiré superlattice, Nat. Phys. **18**, 765 (2022).

[35] K. Kang, T. Li, E. Sohn, J. Shan, and K. F. Mak, Nonlinear anomalous Hall effect in few-layer WTe2, Nat. Mater. **18**, 324 (2019).

[36] M. Suárez-Rodríguez et al., Odd Nonlinear Conductivity under Spatial Inversion in Chiral Tellurium, Phys. Rev. Lett. **132**, (2024).

[37] P. He, H. Isobe, D. Zhu, C.-H. Hsu, L. Fu, and H. Yang, Quantum frequency doubling in the topological insulator Bi2Se3, Nat. Commun. **12**, 698 (2021).

[38] Y. Cao, V. Fatemi, S. Fang, K. Watanabe, T. Taniguchi, E. Kaxiras, and P. Jarillo-Herrero, Unconventional superconductivity in magic-angle graphene superlattices, Nature **556**, 43 (2018).

[39] K. Roy, T. Ahmed, H. Dubey, T. P. Sai, R. Kashid, S. Maliakal, K. Hsieh, S. Shamim, and A. Ghosh, Number-Resolved Single-Photon Detection with Ultralow Noise van der Waals Hybrid, Adv. Mater. **30**, 1704412 (2018).

[40] T. Ahmed, K. Roy, S. Kakkar, A. Pradhan, and A. Ghosh, Interplay of charge transfer and disorder in optoelectronic response in Graphene/hBN/MoS2 van der Waals heterostructures, 2D Mater. **7**, 025043 (2020).

[41] C. Shen et al., Correlated states in twisted double bilayer graphene, Nat. Phys. **16**, 520 (2020).

[42] Y. Wang, J. Herzog-Arbeitman, G. W. Burg, J. Zhu, K. Watanabe, T. Taniguchi, A. H. MacDonald, B. A. Bernevig, and E. Tutuc, Bulk and edge properties of twisted double bilayer graphene, Nat. Phys. **18**, 1 (2022).

[43] *Phys. Rev. B 99, 235406 (2019) - Band Structure and Topological Properties of Twisted Double Bilayer Graphene*, https://journals.aps.org/prb/abstract/10.1103/PhysRevB.99.235406.

[44] Y. Cao et al., Correlated insulator behaviour at half-filling in magic-angle graphene superlattices, Nature **556**, 80 (2018).

[45] F. K. de Vries et al., Kagome Quantum Oscillations in Graphene Superlattices, Nano Lett. **24**, 601 (2024).

[46] M. Suárez-Rodríguez, B. Martín-García, W. Skowroński, K. Staszek, F. Calavalle, A. Fert, M. Gobbi, F. Casanova, and L. E. Hueso, Microscale Chiral Rectennas for Energy Harvesting, Adv. Mater. **36**, 2400729 (2024).